\documentclass[a4paper,twoside]{article}
 \usepackage{amssymb}
\newcommand{\Ref}[1]{(\ref{#1})}
\newcommand{\eqa}{\begin{eqnarray}}
\newcommand{\neqa}{\end{eqnarray}}
\newcommand{\equ}{\begin{equation}}
\newcommand{\nequ}{\end{equation}}
\newcommand{\no}{\nonumber\\}
\def\f{\frac}
\newcommand{\p}{\partial}
\newcommand{\n}{\nabla}

\let\eps=\epsilon
\newcommand{\scr}{\scriptscriptstyle}

\begin{document}

\title{\bf{From 3-geometry transition amplitudes\\ to graviton states}}

\author{ {\bf{Federico Mattei, 
Carlo Rovelli, Simone Speziale, Massimo Testa}\footnote{federico.mattei@roma1.infn.it, 
rovelli@cpt.univ-mrs.fr, simone.speziale@roma1.infn.it, massimo.testa@roma1.infn.it} } 
\\[1mm] 
\em\small{Dipartimento di Fisica
dell'Universit\`a ``La Sapienza", and INFN Sez.\,Roma1, I-00185 Roma,
EU}\\[-1mm] \em\small{Centre de Physique
Th\'eorique de Luminy, Universit\'e de la M\'editerran\'ee, F-13288
Marseille, EU}}

\date{\small\today}
\maketitle
\begin{abstract}
\noindent  In various background independent approaches, quantum gravity is defined in terms of a field propagation kernel: a sum over paths interpreted as a transition amplitude between 3--geometries,  expected to project quantum states of the geometry on the solutions of the Wheeler--DeWitt equation.   We study the relation between this formalism and conventional quantum field theory methods.  We consider the propagation kernel of 4d Lorentzian general relativity in the temporal gauge, defined by a conventional formal Feynman path integral, gauge fixed \`a la Fadeev--Popov.  If space is compact, this turns out to depend only on the initial and final 3--geometries, while in the asymptotically flat case it depends also on the asymptotic proper time.  We compute the explicit form of this kernel at first order around flat space, and show that it projects on the solutions of all quantum constraints, including the Wheeler--DeWitt equation, and yields the correct vacuum and $n$--graviton states.   We also illustrate how the Newtonian interaction is coded into the propagation kernel, a key open issue in the spinfoam approach. 
\end{abstract}

\vskip.5cm

\section{Introduction}

The tentative quantum theories of gravity that are currently better developed, such as for instance strings and loops, are very different from one another in their assumptions and in the formalism utilized.   In particular, the relation between the background--independent methods used in canonical quantum gravity and in the spinfoam formalism, and conventional perturbative quantum field theory  (QFT), is far from transparent \cite{r}.   Besides clouding the communication between research communities, these differences hinder the clarification of a number of technical and conceptual problems.  For instance, there are open questions concerning the physical interpretation of the spinfoam formalism  \cite{Perez}, in particular concerning its low--energy limit, the comparison with quantities computed in the standard QFT perturbative expansion, and the derivation of the Newtonian interaction.  

Here we contribute to the effort of bridging between different languages, by studying the field propagator, or Feynman  propagation kernel, or Schr\"odinger functional, in 4d Lorentzian general relativity (GR).   The  field propagator (and its relativistic extension \cite{r}) is not often utilized in conventional QFT (but see \cite{Symanzik:1981wd} and \cite{Luscher:1992an}), but it is a central object in background--independent quantum gravity \cite{misner,mtw,hawk,teit1}.    Here we analyze this object starting from a conventional path integral quantization of GR and \mbox{--neglecting} its non--renormalizability-- in a perturbative expansion of this integral.  On the one hand, this provides a clean interpretation of a basic non--perturbative tool  in terms of conventional and  well understood quantum field theoretical quantities.  On the other hand, this provides the precise expression of the low--energy limit of the field propagator, to which the non--perturbative one must be compared.  In other words, we study the explicit relation between the 3--geometry to 3--geometry transition amplitude and the graviton--state language. 

The key to bridge between a conventional path integral formulation and the non--perturbative framework is the use of the temporal gauge, with a careful implementation of the Faddeev--Popov (FP) gauge--fixing.  Conventional perturbation theory is usually studied in covariant gauges such as the Lorentz gauge in Yang--Mills (YM) theory or the harmonic gauge in GR \cite{Veltman}, but the temporal gauge is naturally closer to the Hamiltonian formalism.  The propagation kernel of YM theory in the temporal gauge is well understood \cite{rt}.  It  can be computed as a Feynman path integral for finite time, with fixed initial and final field configurations.  In the temporal gauge there are no dynamical ghosts, and the kernel has a clear physical interpretation: it gives the matrix element of the evolution operator between eigenstates of the YM connection. Namely, the transition amplitude between the states defined on the boundaries. But it is also a projector on the physical states, which satisfy the YM quantum constraints.  At the zero'th order in perturbation theory, the kernel can be explicitly computed with a gaussian integration in the Euclidean regime, and it nicely codes the form of the perturbative vacuum as well as all the $n$--particle states \cite{rt,vacuum,filippo}.  Furthermore, it is straightforward to express the $n$--point functions in terms of it.  In spite of key differences, the structure of GR is similar to a YM theory in many respects.   We can apply to GR the techniques used for YM theory, and, in particular, study the propagation kernel of quantum GR in the temporal gauge. This is what we do in this paper. Notice that the use of non--covariant gauges has been considered
in the literature (see for instance \cite{Delbourgo:1981gv}).

First, we consider the formal path integral that defines the propagation kernel in the temporal gauge, and study its properties.   This object has been considered in the literature (see \cite{teit1, teit2} and references therein),
though with different techniques.
The expression is formal because the measure is unknown, and the usual perturbative definition is not viable because
the perturbative expansion on a background  is non--renormalizable. We do not consider background--independent definitions of the path integral, such as the spinfoam one, because our interest here is the interpretation and the low energy limit of the kernel, not its ultraviolet divergences. We consider the two possibilities of compact and asymptotically--flat space
 \cite{teit3}. We carefully discuss the FP gauge fixing and in this context we identify the integration implementing the quantum constraints as the one on the FP gauge parameters.  We discuss how the propagation kernel turns out to be independent from the coordinate time in both cases --a feature that drastically distinguishes GR from YM theory--, but to depend on the asymptotic proper--time in the second case.   We use the conventional metric formalism for GR.  
The expression we obtain is formal, and we do not discuss topological and ultraviolet aspects.

Second, we consider the zero'th order term in the perturbative expansion of the integral on a flat background, and we compute the propagation kernel explicitly.   We show that it projects on the solutions of all the constraints, and that it correctly codes the perturbative GR vacuum state \cite{Kuchar,har} and the $n$--particle states. This provides an explicit bridge between 3--geometry transition amplitudes and  perturbative graviton states. The non--perturbative boundary amplitudes of the spinfoam formalism must reduce to this expression for boundary metrics close to flat space.\footnote{For recent applications of this idea, see \cite{prop}.}

Finally, we couple an external matter source to the theory, and derive the expression for the energy of the field in the presence of matter.  This expression codes the Newtonian interaction. Our hope is that this could open the way for extracting the Newtonian interaction from the spinfoam amplitudes, hence providing a key missing check  of their physical viability. 

We fix the speed of light and the Newton and Plank constants by $c=16\pi G = \hbar = 1$. Greek indices range from 0 to 3; latin indices from 1 to 3.\  We use coordinates $x^\mu = (t, \vec{x})$. $A_\mu(x)$ is the YM field, $A_i(\vec x)$ the vector potential; $g_{\mu\nu}(x)$ is the gravitational field and the spacetime metric; while $g_{ij}(\vec x)$ is the metric of a spacelike surface.  (Indices are intended in Geroch's abstract index notation: $A_i$ means $\vec A$, and so on.)

\subsection{YM propagation kernel in temporal gauge}\label{sectionYM}

Before turning to GR, we recall the properties of the field propagation kernel in YM theory.  As we shall see, the gravitational case will present substantial analogies, as well as key differences.  Formally, the propagation kernel is given by the functional integral over the configurations of the YM field defined on the spacetime region bounded by the initial and final surfaces $t=0$ and $t=T$, and restricted to given initial and final configurations $A_\mu(\vec x, 0)=A'_\mu(\vec x)$ and  $A_\mu(\vec x, T)=A''_\mu(\vec x)$
\begin{equation}
\label{1ym}
W[A_{\mu}',A_{\mu}'',T]=\int _{A_{\mu}'' \atop A_{\mu}'} \mathcal{D} A_{\mu} \ e^{i{S[A_{\mu}]}}, 
\end{equation}
where 
\begin{equation}
\label{2ym}
S[A_\mu]=\int_0^T dt \int  d^3x \   \mathcal{L}_{\rm YM} 
\end{equation}
and $ \mathcal{L}_{\rm YM}$ is the YM lagrangian. The integral \Ref{1ym} contains an  infinity due to the integration over the group of the gauge trasformations $A\to A^\Lambda$, where $\Lambda(\vec x, t)$ is the gauge parameter.  We fix this by gauge fixing \Ref{1ym} \`a la FP
in the temporal gauge $A_0=0$ \cite{rt}.  That is, we insert in \Ref{1ym} the identity
\begin{equation}
\label{3ym}
1=\Delta_{\rm FP} (A_{\mu}) \int {\cal D}\Lambda \, \delta(A^\Lambda_{0}).
\end{equation} 
The FP determinant $\Delta_{\rm FP}$ is actually a constant, 
because the YM gauge transformations do not mix the different components of the 4--vector $A_\mu$, hence $\Delta_{\rm FP}$ can only depend on $A_0$; but this, in turn, is fixed to be zero by the $\delta$--function appearing in the integral. The propagation kernel between boundary values of the 3d YM connection $A_i$ is thus given by  
\begin{equation}
\label{11ym}
W[A_{i}',A_{i}'',T]=
\Delta_{\rm FP}
\int _{A_{i}'' \atop A_{i}'}  \mathcal{D} A_{\mu}
 \int {\cal D}\Lambda \, \delta(A^\Lambda_{0})
\ e^{i{S[A_{\mu}]}}.
\end{equation}
Changing the order of the two integrations, changing variables $A_\mu\to A_\mu^\Lambda$ and integrating over $A_0$, we obtain 
\begin{equation}
\label{12ym}
W[A_{i}',A_{i}'',T]=
\Delta_{\rm FP}
 \int {\cal D}\Lambda  
\int _{A_{i}''{}^{\Lambda (T)} \atop A_{i}'{}^{\Lambda (0)}}  \mathcal{D} A_{i}
\ e^{i{S[A_{i};A_0=0]}}.
\end{equation}
The only components of the integration variable $\Lambda$ entering the integrand are its values at $t=0$ and $t=T$. We can therefore drop the bulk integration on $\Lambda(\vec x, t), 0>t>T$, discarding a trivial infinity.   Furthermore, since $ \mathcal{D} A_{i}$ and $S[A_{i},A_0=0]$ are invariant under time independent gauge trasformations, the $\mathcal{D} A_{i}$ one of the two remaining integrals on  $\Lambda(\vec x, 0)$ and  $\Lambda(\vec x, T)$ is redundand. Dropping the second, we have therefore
\begin{equation}
\label{4ym}
W[A'_{i},A_{i}'',T] =
 \int {\cal D} \lambda \  \tilde W[A'_i{}^{\lambda},A''_i,T]
\end{equation}
where
\begin{equation}
\label{5ym}
\tilde W[A'_{i},A''_{i},T]=\Delta_{\rm FP} 
\int _{A_{i}'' \atop A_{i}'} \mathcal{D} A_{i}
 \ e^{i{S[A_{i},A_0=0]}},
\end{equation}
and $\lambda(\vec x)$ is the gauge parameter of the residual time--independent gauge trasformations $A_i\to A_i^\lambda$ of the $A_0=0$ gauge. The propagator  $\tilde W[A'_{i},A''_{i},T]$  is invariant under \emph{simultaneous} gauge trasformations on the two boundaries. 
The integral over $\lambda$ in \Ref{4ym} makes $W[A'_{i},A''_{i},T]$ invariant under \emph{independent} gauge trasformations on the two boundaries.   
 
The field propagator is the matrix element of the evolution operator between eigenstates of the field operator $A_i$ 
\begin{equation}
\label{6ym}
W[A'_{i},A_{i}'',T]= \langle  A_{i}''  | e^{-iHT} | A'_{i}  \rangle,
\end{equation} 
where $H$ is the hamiltonian; up to the difficulties in defining fixed--time operators in an  interacting QFT (see \cite{Symanzik:1981wd}), it can be interpreted as the Feynman probability amplitude of having the field configuration $A''_i$ at time $T$, given  the field configuration $A'_{i}$ at time $0$.  Equivalently, it time--propagates the quantum state in the Schr\"odinger functional representation of the quantum field theory 
\begin{equation}
\label{7ym}
\Psi_{t+T}[A_i] = \int  \mathcal{D} A'_i\  W[A_{i},A'_{i},T] \  \Psi_t[A'_i].
\end{equation}
However, notice that, because of its gauge--invariance, any state obtained by propagating with the kernel is invariant under gauge trasformations of $A_i$.  Therefore the kernel is also a projector on the gauge--invariant, or ``physical", states, which satisfy $\Psi[A_i]=\Psi[A^\lambda_i]$. That is, the states that satisfy the Gauss--law quantum constraint.   
Thus, the ${\cal D}{A_i}$ integral \Ref{5ym} takes care of the gauge--variant dynamics, and the ${\cal D}\lambda$ integral  \Ref{4ym}  imposes gauge invariance. 

If $\Psi_n(A_i)$ is a basis of gauge--invariant physical states that diagonalize the energy, \Ref{6ym} implies
\begin{equation}
\label{6YM}
W[A_{i},A'_{i},T] =
\sum_n  e^{-iE_nT}\ 
 \overline{ \Psi_n[A'_i]} 
  \Psi_n[A_i].
\end{equation}
In particular, once subtracted the zero-point energy, we can read out the form of the vacuum state from the propagator 
\begin{equation}
\label{vuoto}
\int_{-\infty} ^{+ \infty} dT\  W[A_i,A'_i=0,T] = const\   \Psi_0[A_i].
\end{equation}
Thus, the temporal--gauge propagation kernel nicely bridges between the functional integral formalism and the hamiltonian one. 

The propagation kernel $W[A'_{i},A_{i}'',T]$ can be computed explicitely order by order in  perturbation theory, by a gaussian integration in the Euclidean regime.   For this to be well defined, we need to assume appropriate boundary conditions for the field at spacial infinity. In particular, we assume that the field vanishes at spacial infinity, and therefore so has to do the gauge parameter.  In the lowest order (or exactly in the Maxwell case) the propagator kernel turns out to be \cite{rt}
\eqa 
W[A_{i}',A_{i}'',T] &=& \mathcal{N} (T) \; 
e^{\textstyle\frac{i}{2} 
{\displaystyle \int} \frac{d^3p}{(2\pi)^3} p
\frac{(|A'{}^T|^2+|A''{}^T|^2) 
\cos{p T} - 2A'{}^T\cdot A''{}^T}{\sin{p T}}}. 
\neqa 
Here $p=|p|=\sqrt{p_ip^i}$, the Fourier trasform of the potential is 
$A_j(\vec p)=\int d^3x\, e^{i\vec p\cdot \vec x} A_j(\vec x)$ and its transverse component is defined as $A^T_i(\vec p)\equiv D_i{}^j(\vec p) A_j(\vec p)$, where 
\begin{equation}
\label{D}
D_{ij}= \delta_i^j-\frac{p_ip_j}{p^2}.
\end{equation}
We can read out from this kernel all the free $n$--particle states with momenta $p_\alpha$, energy $E_\alpha=p^0_\alpha$ and polarizations $\epsilon_\alpha$, $\alpha=0,...,n$, from the expression
\equ\label{ZYM2}
W[{A}_i, A'_i, T]=\frac{1}{n!} \sum_{n=0} ^\infty\;
\sum_{\epsilon_1...\epsilon_n}
\int \frac{d^3p_1}{(2\pi)^3}\ldots \frac{d^3p_n}{(2\pi)^3} 
\;e^{-i \sum_{\alpha=1}^n E_\alpha T} 
\;\overline{\psi_{p_1\epsilon_1,...,p_n\epsilon_n}[{A}_i]} \;\psi_{p_1\epsilon_1,...,p_n\epsilon_n}[A'_i].
\nequ

Let us now turn to general relativity.

\section{Propagation kernel in general relativity}\label{sectionPI}

We foliate spacetime with a family of 3d surfaces $\Sigma_t$, with $t\in R$, and 
and focus on the region $t\in[0,T]$. We fix initial and final positive definite metrics $g_{ij}'$ on $\Sigma_0$ and $g_{ij}''$ on $\Sigma_T$ as boundary data, and 
we want to describe the quantum dynamics of the gravitational field in terms of the functional integral
\begin{equation}
\label{1}
W[g_{ij}',g_{ij}'',T]=\int _{g_{ij}'' \atop g_{ij}'} \mathcal{D} g_{\mu\nu} \ e^{i{S[g_{\mu\nu}]}}
\end{equation}
over the 4d spacetime metrics inducing the given 3d metrics on the two boundaries. 
The action of the gravitational field is
\begin{equation}
\label{2}
S[g_{\mu\nu}]=  \int_0^T dt \int _{\Sigma_t} d^3x \  \sqrt{-g}\,g^{\mu\nu}R_{\mu\nu}\ +\ \int_{\Sigma_0\cup\Sigma_T}  d^3x \ \mathcal{K} \  \equiv \    \int_0^T dt \int _{\Sigma_t} d^3x \ 
 \mathcal{L}. 
\end{equation}
Here $R_{\mu\nu}$ is the Ricci tensor and $g$ the determinant of the metric. $\mathcal{K}$
is the extrinsic curvature of the boundary surfaces. The presence of the boundary term, sometimes called the  {\it Gibbons--Hawking term},  is needed in order to have only first--order time derivatives in the action $\mathcal{L}$, so that the convolution property of the propagation kernel is guaranteed \cite{gh}.  

There are various sources of infinities in \Ref{1}.  First, there are ultraviolet divergences. As discussed in the introduction,  we disregard them here, under the assumption that an appropriate non--perturbative definition of the integral, such as in the spinfoam formalism, could take care of them.    

Second, we must be sure that a sufficient number of boundary conditions are fixed. In general,
we may reasonably demand that a classical solution is uniquely selected by these conditions by minimizing the action.  We focus on two cases:  (i) The compact case, in which the Riemaniann manifolds $(\Sigma_0, g_{ij}')$ and  $(\Sigma_T, g_{ij}'')$ have finite volume and no boundary. In this case, initial and final 3--metrics may determine a unique classical solution. The {\it thin sandwich conjecture} \cite{mtw,bsw},  indeed, states that, at least for small times, generically there is only one spacetime metric between two given spacelike metrics.   Exceptions are known, generally characterized by pathologies such as singularities \cite{bf}.  In the following we restrict to the cases where the conjecture holds. Amongst the exceptions is the remarkable case of flat space, as emphasized in \cite{Peres}.  (ii) The asymptotically flat case, in which $\Sigma$ is homeomorphic to ${{\mathbb R}}^3$ and $g_{ij}'$ and  $g_{ij}''$ appropriately converge to $\delta_{ij}$ at infinity. In this case, for \Ref{1} to be well defined, we demand $g_{\mu\nu}$ to converge asymptotically to the Minkowski metric $\eta_{\mu\nu}$ for all $t$. The consequences for the uniqueness of the solution will be discussed in details below.\footnote{\ It has been recently suggested that the most interesting case for nonperturbative quantum gravity is when $\Sigma_t$ has boundaries \cite{noi}.  For a discussion of the classical solutions in this case, see \cite{Nagy}.}

Third, the invariance under diffeomorphisms of GR makes the integral \Ref{1} infinite. This situation is analogous to the YM case, and can be cured with a gauge--fixing, as we do below. 

\subsection{Gauge--fixing}\label{sectionGf}

General relativity is invariant under diffeomorphisms, namely under the pull back 
$g\to g^\xi=\xi_*g$ of the gravitational field by a map $\xi\!\!:M \to M$ from the spacetime to itself. These are  the gauge transformations of GR.\footnote{\ They should not be confused with the freedom of chosing coordinates on $M$: once coordinates are fixed, the integral \Ref{1} still gets contributions from distinct fields $g$ and $g^\xi$.} Explicitly, $g^\xi$ is defined by 
\equ\label{gcsi1}
g_{\mu\nu}(x) =\f{\p \xi^\rho(x)}{\p x^\mu} \f{\p \xi^\sigma(x)}{\p x^\nu}\; g^\xi_{\rho\sigma}(\xi(x))
\nequ
where $\xi:x^\mu\mapsto\xi^\mu(x)$. \big(Or $g^\xi_{\mu\nu}(x) =\f{\p \phi^\rho}{\p x^\mu} \f{\p \phi^\sigma}{\p x^\nu}\; g_{\rho\sigma}(\phi(x))$ where $\phi=\xi^{-1}$, a form we'll use later on.\big)    For the action \Ref{2} to be invariant, $\xi$ must not change the boundaries of the spacetime region considered, that is 
\equ\label{csisulbordo}
\xi^0(0,\vec{x})=0, \qquad \xi^0(T,\vec{x})=T.
\nequ
Because of this gauge invariance, the integral (\ref{1}) has an infinite contribution from the integration over the gauge group.   We take care of this as we did for YM, by introducing a (non--covariant)
gauge--fixing. The GR analogue of the temporal gauge is known as ``gaussian normal coordinates", or the ``Lapse=1, Shift=0", or ``proper--time" gauge
\equ\label{gaugefixing}
g_{00}=-1, \qquad g_{0i}=0.
\nequ 
As for YM, this is not a complete gauge--fixing, but we expect that additional gauge--fixing is not required in the path integral.  In the linearized case we shall explicitly see that the remaining part of the gauge is taken care by the integration over the gauge parameters.  Thus, we gauge--fix the path integral by inserting in \Ref{1} the FP identity
\begin{equation}
\label{33}
1=\Delta_{\rm FP} [g_{\mu\nu}] \int {\cal D}\xi  \  \delta(g^\xi_{00}+1) \, \delta(g^\xi_{0i}).
\end{equation}
${\cal D}\xi$ is a formal measure over the group of the 4d diffeomorphisms.  We can see here a first difference with YM theory: the GR gauge transformations mix the different components of $g_{\mu\nu}$, hence the FP factor $\Delta_{\rm FP}$ is not a constant anymore.  Since the four $g_{0\mu}$ are fixed by the $\delta$--functions, it will depend only on the spacial components of the metric.  

The integration \Ref{33} is over all $\xi^\mu(\vec x, t)$ with $t\in[0,T]$, including the boundaries, and it is \emph{not} restricted by \Ref{csisulbordo}, therefore it includes $\xi$ that change the action
\Ref{2}. To understand this delicate point, observe that the FP integral must include sufficient gauge trasformations for transforming \emph{any} field to one satisfying \Ref{gaugefixing}.We cannot fix $g_{00}=-1$, and also the coordinate time between initial and final surface. This would amount to discard all four--metrics yielding a proper time between the two surfaces different from $T$;  but these field configurations do contribute to \Ref{1} and cannot be discarded.   In other words, we \emph{cannot} gauge transform all fields contributing to the integral to the gauge \Ref{gaugefixing} without changing the action. However, nothing prevents us from transforming them to fields satisfying \Ref{gaugefixing} \emph{and} changing the action when needed. (See also \cite{teit1} on this.)

Introducing (\ref{33}) in (\ref{1}) we obtain
\begin{equation}
\label{4}
W[g_{ij}',g_{ij}'', T]= \int _{g_{ij}''  \atop g_{ij}'} \mathcal{D} g_{\mu\nu}\, \Delta_{\rm FP}[g_{ij}]
 \int {\cal D}\xi\ \exp{\left\{i\int d^3x \int_0^T dt \, {\mathcal{L}}[g_{\mu\nu}]\right\} }  \delta(g^\xi_{00}+1) \, \delta(g^\xi_{0i}).
\end{equation}
We now evaluate the integrals over $g_{0\mu}$. As we did in the YM case,
we exchange the order of the integrations and change variables $g_{\mu\nu}\rightarrow  g_{\mu\nu}^\xi$, together
with a change of coordinates $x^\mu\to \xi^\mu(x)$ in the action. As a consequence, the
boundary data will now depend on $\xi$. We perform the integrals over $g_{00}$ and $g_{0i}$, obtaining 
\begin{equation}
\label{51}
W[g_{ij}',g_{ij}'', T]= \int {\cal D}\xi\ \  \tilde W[g_{ij}',g_{ij}'',\xi ,  T]
\end{equation}
where
\begin{equation}
\label{52}
\tilde W[g_{ij}',g_{ij}'',\xi, T]=
\int _{g''_{ij}{}^\xi  \atop g_{ij}'{}^\xi} \mathcal{D} g_{ij} \,\Delta_{\rm FP} [g_{ij}] \exp{\left\{i\int d^3x \int_{\xi^0(\vec x, 0)}^{\xi^0(\vec x, T)} dt \, {\mathcal{L}}[g_{ij}, g_{00}=1, g_{0i}=0, ]\right\} }.
\end{equation}

Let us analyze these expressions. The $\mathcal{D} g_{ij}$  integral in \Ref{52} is over the fields in the spacetime region bounded by the surfaces $\Sigma^\xi_0$ and $\Sigma^\xi_T$, defined respectively by $t=\xi^0(\vec x, 0)$ and $t=\xi^0(\vec x, T)$, having boundary value $g''_{ij}{}^\xi$  on $\Sigma^\xi_0$ and $g_{ij}'{}^\xi$ on $\Sigma^\xi_T$.  Notice that  $g'_{ij}{}^\xi$ depends precisely on $g'_{ij}$ and $\xi$, since, using \Ref{gaugefixing}, the transformation of $g_{ij}$ reads
\equ\label{gcsi}
g^\xi_{ij}(x)=\f{\p \xi^\rho}{\p x^i} \f{\p \xi^\sigma}{\p x^j} g_{\rho\sigma}(\xi(x))=
-\f{\p \xi^0}{\p x^i} \f{\p \xi^0}{\p x^j} +
\f{\p \xi^k}{\p x^i} \f{\p \xi^l}{\p x^j} g_{kl}(\xi(x));
\nequ
that is, the value of $g'_{ij}{}^\xi$ on $\Sigma^\xi_0$ is determined by $g'_{ij}$ on $\Sigma_0$ and by the map $\xi:\Sigma_0\to\Sigma_0^\xi$.
 
Equation \Ref{52} depends only on the \emph{boundary} diffeomorphisms  $\xi_{\rm ini}^\mu(\vec x)=\xi^\mu(\vec x, 0)$ and $\xi_{\rm fin}^\mu(\vec x)=\xi^\mu(\vec x, T)$. 
The integral over the bulk diffeomorphisms depends in general on the boundary surfaces.\footnote{We
thank our referee for pointing this out.} The result of this integration is an appropriate functional
such that the convolution property of the kernel will be satisfied. However, since we are interested
in explicitly computing the linear approximation, this term is not relevant and it will be discarded
in the following.
The boundary gauge transformations play a more subtle role 
than in YM theory. Indeed, notice that we cannot write $\tilde W[g_{ij}',g_{ij}'',\xi, T]$ in the form $\tilde W[g_{ij}'^\xi,g_{ij}''^\xi, T]$ as we did for YM, because $\xi$ affects also the action.  Let us distinguish the temporal boundary diffeomorphisms $\xi^0_{\rm ini}$ and $\xi^0_{\rm fin}$ from the spacial ones, $\xi^i_{\rm ini}$ and $\xi^i_{\rm fin}$. Truly,  \Ref{52} is invariant under a spacial diffeomorphism which acts identically on both boundaries. This fact is analougous to the YM case, and allows us to drop the dependence on $\xi^i$ of one of the boundaries, say  at $t=0$.  On the other hand, this property is not true for temporal diffeomorphisms as emphasized in \cite{teit1}.
Thus we can write, introducing the shorthand notation
${\mathcal{L}}[g_{ij}]\equiv{\mathcal{L}}[g_{ij}, g_{00}=-1, g_{0i}=0]$, 
\begin{equation}
\label{6}
W[g_{ij}',g_{ij}'', T] = \int {\cal D}\xi_{\rm fin}^i\; {\cal D}\xi_{\rm fin}^0\; {\cal D}\xi_{\rm ini}^0  
\int_{g''_{ij}{}^{\xi^\mu}  \atop g_{ij}'{}^{\xi^0}} \mathcal{D} g_{ij} \,\Delta_{\rm FP} [g_{ij}] 
\exp{\left\{i\int d^3x \int_{\xi^0(\vec x, 0)}^{\xi^0(\vec x, T)} dt \, {\cal{L}}[g_{ij}]\right\} } .
\end{equation}
This is the exact expression for the propagation kernel of GR.
We expect the integration over the gauge parameters to implement the constraints of the theory. 
In section 3 below, we show explicitly how this happens in the linearized case. Before that,
however, we discuss a feature of the kernel which is characteristic of GR. 

\subsection{The disappearance of time}

Let us first focus on the case (i) in which $(\Sigma, g_{ij})$ is a compact space with finite volume, and consider the ${\cal D}\xi_{\rm fin}^0$ integration in \Ref{6}.  This is an integration over all possible coordinate positions of the final surface. This ensemble is the same whatever is $T$, hence the right hand side of \Ref{6} does not depend on $T$.  Therefore in this case the propagation kernel of GR is independent from $T$, and we can simply write 
\begin{equation}
\label{66}
W[g_{ij}',g_{ij}''] = \int {\cal D}\xi_{\rm fin}^i\; {\cal D}\xi_{\rm fin}^0\; {\cal D}\xi_{\rm ini}^0  
\int_{g''_{ij}{}^{\xi^\mu}  \atop g_{ij}'{}^{\xi^0}} \mathcal{D} g_{ij} \,\Delta_{\rm FP} [g_{ij}] 
\exp{\left\{i\int d^3x \int_{\xi^0(\vec x, 0)}^{\xi^0(\vec x, 1)} dt \, {\cal{L}}[g_{ij}]\right\} }. 
\end{equation}
This is what we mean by disappearance of time. Notice however that a proper time is (generically) determined by the initial and final 3--metrics themselves. In fact, the integral \Ref{66} is likely to be picked on the \emph{classical} solution $g_{\mu\nu}$ bounded by $g_{ij}'$ and $g_{ij}''$. But
recall that in classical GR the initial and final 3--metrics are expected to generically determine a classical time lapse between them.  To see how this can happen,  consider the theory in the partial gauge--fixing $g_{0i}=0$. The six evolution equations for $g_{ij}(\vec x, t)$ are second order equations that we expect to generically admit a solution for given initial and final data at $t=0$ and $t=1$. The time--time component of the Einstein equations --the scalar constraint-- can then be written in the form 
\begin{equation}
\label{tt}
\left(g^{ik}g^{jl}-\f{1}{2}g^{ij}g^{kl}\right) \f{\p g_{ij}}{\p t} \f{\p g_{kl}}{\p t}-g_{00}\; 
\det{g_{ij}}\; R[g_{ij}]=0,
\end{equation}
where $R$ is the Ricci scalar of the 3--metric $g_{ij}$.
Once a solution $g_{ij}(\vec x, t)$ is given, this equation can be immediately solved algebraically for $g_{00}(\vec x, t)$ and therefore it determine the physical proper time 
\begin{equation}
\label{T(x)}
T(\vec x)
= \int_0^1  \sqrt{g_{00}(\vec x, t)}\ dt
= \int_0^1  \sqrt{\f{\left(g^{ik}g^{jl}-\f{1}{2}g^{ij}g^{kl}\right)\f{\p g_{ij}}{\p t} \f{\p g_{kl}}{\p t}}{\det{g_{ij}} \ R[g_{ij}]}}\ dt
\end{equation}
between initial and final surface, along the $\vec x= const$ lines, which in this gauge are geodesics normal to the initial surface.  Hence in general this proper time is determined by the initial and final 3--geometries.  On the other hand, notice that as an equation for $g_{00}$, \Ref{tt} becomes indeterminate when $R[g_{ij}]=0$, and in particular on flat space.    

The disappearance of the time coordinate in \Ref{66}, and in general in the transition amplitudes of quantum gravity, and its physical interpretation, have been amply discussed in the literature. Its physical meaning is that GR does not describe physical evolution with respect to a time variable representing an external clock, but rather the relative evolution of an ensemble of partial observables. See for instance \cite{r} for a detailed discussion, and the references therein. 

The invariance of GR under coordinate time reparametrization implies that the hamiltonian $H$ in \Ref{6ym} vanishes, and therefore we expect to have
\equ
\label{Zgr}
    W[g_{ij}',g_{ij}''] = \sum_n\; \overline\Psi{}_n[g_{ij}'']\; \Psi_n[g_{ij}']
\nequ
instead of \Ref{6YM}, where again $\Psi_n[g_{ij}]$ is a complete basis of physical states, satisfying all the Dirac constraints of GR, including, in particular, the Wheeler--DeWitt equation, which codes the quantum dynamics of the theory in the hamiltonian framework.  Equation \Ref{Zgr} indicates that  $W[g_{ij}',g_{ij}''] $ is the kernel of a projector on the physical states of the theory \cite{P}.
Roughly speaking, the ${\cal D}\xi^i$ integration implements the invariance under spacial diffeomorphisms, while the  ${\cal D}\xi^0$ integrations project on the solutions of the Wheeler--DeWitt equation. We will see that this is indeed the case in the linear approximation. 

An explicit perturbative computation of the propagation kernel in the compact case would be very interesting.  The problem of expanding around a flat background in the compact case is that this is precisely one of the degenerate cases where the thin--sandwhich conjecture fails;  essentially because, as we have seen, \Ref{tt} fails to determine $g_{00}$ in this case.  However, 
this can probably be simply circumvented, for instance by adding a small cosmological constant. 
We leave this issue for further developements, and we turn, instead, to the asymptotically flat case, where the explicit computation of the propagation kernel can be performed in a more straightforward fashion.

Thus, consider the case (ii) in which $(\Sigma, g_{ij})$ is asymptically flat.  In this case,
again \Ref{tt} does not determine $T$. However, the argument above for the disappearance
of time fails, for the following reason. As we have already mentioned, for the path integral to be well defined we must require flat (\emph{i.e.} $g_{ij}=\delta_{ij}$) boundary conditions at spacial infinity for $g_{ij}(\vec x, t)$.  In order to be well defined on this space of fields, the gauge trasformations must vanish at infinity accordingly, as they are required to do in the YM case.  Therefore $\xi^0(\vec x, T)$ must converge to $\xi^0(\vec x, T)=T$ for large $\vec x$, and the right hand side of \Ref{6} is not independent from $T$. In fact, what is physically relevant is obviously not the coordinate time, but rather the proper time separation between initial and final surfaces at infinity.  That is, in the asymptotically flat case, the GR field propagator depends on the asymptotic proper time at infinity.\footnote{\ A suggestive way of understanding the presence of asymptotic time in the asymptoticaly flat case, based on the boundary interpretation of the time evolution developed in \cite{r,noi} is the following.  According to \cite{r,noi}, the observable proper time in GR can be determined by the boundary value of the propagation kernel of a finite spacetime region, namely a 4d ball. In the limit in which the spacial dimensions of the region go to infinity, the boundary proper time converges to the asymptotic proper time.}

 \section{Linearized Theory}\label{sectionLin}

We now focus on the asymptotic case, write 
\eqa\label{etah}
g_{\mu\nu}(x)=\eta_{\mu\nu}+h_{\mu\nu}(x)
\neqa
 and consider the field $h_{\mu\nu}(x)$ as a perturbation.  Since we restrict to small $h_{\mu\nu}$ fields, we restrict, accordingly, to small diffeomorphisms, that preserve the form \Ref{etah}. The gauge condition \Ref{gaugefixing} implies $h_{0\mu}(x)=0$. In this gauge, the action \Ref{2} reads
 \eqa\label{actionLin}
&&\int d^3x \int_0^T dt \ {\cal L}= \int d^3x \int_0^T dt\;\Bigg\{
{\partial_{i}\partial_{j}h^{ij}}- \nabla^2{h} + \nonumber\\ &&+
\frac{1}{4}\Big[ \partial_\alpha h^i_i \partial^\alpha h^i_i - \partial^{\alpha}h^{ij} \partial_\alpha h_{ij}+ \partial^k h^{ij} \partial_j h_{ik} + \partial^k h^{ij} \partial_i h_{jk} - \partial_i h^{ij} \partial_j h - \partial^j h_{ij} \partial^i h \Big]\Bigg\}.
\neqa
Indices are now raised and lowered with the Minkowski metric and the 3d Euclidean metric.  It is convenient to separate the four Poincar\'e--invariant components of the linearized field $h_{ij}$: 
the spin--zero trace of the transverse components $h_0$, 
the spin--one and spin--zero longitudinal components $h_i = h_i^{\rm \scriptscriptstyle T} + \frac{p_i}{p} h^{\rm \scriptscriptstyle L}$,  and the spin--two traceless and transverse components $h^{\rm \scriptscriptstyle TT}_{ij}$.   In Fourier space, this decomposition is
\begin{equation} \label{poincare} h_{ij}=-2h_0D_{ij} +h^{\rm\scr T}_{i}\frac{p_{j}}{p}
+h^{\rm\scr T}_{j}\frac{p_{i}}{p}+h^{\rm\scr L}\frac{p_ip_j}{p^2}
+h^{\rm \scriptscriptstyle TT}_{ij}
\end{equation}
where $D_{ij}$ is the projector on the transverse modes given in \Ref{D} and 
\begin{equation}
\label{spin}
p^{i}h^{\rm \scriptscriptstyle TT}_{ij}=0, \qquad \delta^{ij}h^{\rm \scriptscriptstyle TT}_{ij}=0
, \qquad p^ih^T_i=0.
\end{equation}
See Appendix C for more details. 

The action is invariant under infinitesimal spatial diffeomorphisms, $x^i\rightarrow 
\xi^i(x)\simeq x^i+\eps^i(x)$, which induce the transformation 
$h_{ij}(x)\rightarrow h^\epsilon{}_{ij}(x)= h_{ij}(x)+\p_i \eps_j(x) + \p_j \eps_i(x)$
on the linearized field. Or
\begin{equation}\label{lindiff}h_{i}(x)\rightarrow h^\epsilon{}_{i}(x)= h_{i}(x)+ \eps_i(x).
\end{equation}
Note, from this last expression, that $\epsilon_i(x)$ is of the same order of $h_{ij}(x)$.

As mentioned, the role of the temporal infinitesimal diffeomorphism is subtle: the lagrangian
is insensitive to them, but they affect the extrema in the $t$ integral in the 
action.\footnote{\ Condition \Ref{csisulbordo} for the invariance of the action now reads
$\eps^0(\vec{x},0)=0, \ \eps^0(\vec{x},T)=0.$}
The term ${\partial_{i}\partial_{j}h^{ij}}- \nabla^2{h}=-2\nabla^2{h_0}$ 
appearing in the lagrangian is a boundary term, and does not contribute to the equations of motion, but it does contribute to the integral. It is the linear approximation to the left hand side of the time--time component of the Einstein equations \Ref{tt}, namely of the GR scalar constraint.

The boundary data $h_{ij}(0,\vec{x})=h'_{ij}(\vec{x})$ and $h_{ij}(T,\vec{x})=h''_{ij}(\vec{x}),
$  vanish at spacial infinity. Hence \Ref{6} becomes
\begin{equation}
W[h_{ij}',h_{ij}'',T]= \int {\cal D}\epsilon^i_{\rm fin}\; 
{\cal D}\epsilon_{\rm fin}^0\; {\cal D} \epsilon_{\rm ini}^0  \int _{h_{ij}''{}^{\eps^i} \atop h_{ij}'} \mathcal{D} h_{ij} 
\;\Delta_{\rm FP}[h_{ij}]\ 
e^{i\int d^3x \int_{0+\epsilon^0(0, \vec x)}^{T+\epsilon^0(T, \vec x)} dt \, {\mathcal{L}}[h_{ij},h_{0\mu}=0] } .
\label{11}
\end{equation}
The ${\cal D}\epsilon^i_{\rm fin}$ integration implements the invariance \Ref{lindiff} separately on initial and final data.  It thus makes $W[h_{ij}',h_{ij}'',T]$ independent from the spin--one and longitudinal spin--zero longitudinal components $h_{i}'$ and $h_{i}''$. 
How about the integrations over the $\eps^0$'s?
Since the background metric is static, $\eps^0$ enters only the action's boundaries.\footnote{
\ On a generic background $g^0_{ij}$, the transformation under diffeomorphisms of the perturbation
is, at first order, $h_{ij}(x)\rightarrow h^\epsilon{}_{ij}(x)= h_{ij}(x)+\p_i \eps_j(x) + \p_j \eps_i(x) +
\eps^\mu\p_\mu g^0_{ij}(x)$. Therefore, $\eps^0$ would enter the boundary data.} 
Taylor expanding the action we have 
\begin{eqnarray}
\nonumber
&& \int d^3x \int_{0+\epsilon^0(\vec x,0)}^{T+\epsilon^0(\vec x,T)} dt \; \mathcal{L}[h_{ij},h_{0\mu}=0]
\simeq \\\nonumber 
&& \simeq \frac{1}{4}\int\! d^3x \int_{0}^{T}\!\! dt 
\Bigg\{  \partial_\alpha h \partial^\alpha h - \partial^{\alpha}h^{ij} \partial_\alpha h_{ij}
+ \partial^k h^{ij} \partial_j h_{ik} + \partial^k h^{ij} \partial_i h_{jk} - \partial_i h^{ij} \partial_j h - \partial^j h_{ij} \partial^i h \Bigg\}\\ \label{12}
&&+\int d^3x \; \Bigg\{\epsilon^0(T, \vec x) \, \left( \partial_{i}\partial_{j}h_{ij} - \nabla^2 h
 \right){\Big|_{t=T}}-
\epsilon^0(0, \vec x) \, \left( \partial_{i}\partial_{j}h_{ij} - \nabla^2 h \right){\Big|_{t=0}}\Bigg\}.
\end{eqnarray}
Thus, the integration over the $\eps^0$'s gives two $\delta$--functions of the linear term
${\partial_{i}\partial_{j}h^{ij}}- \nabla{h}$ on the boundary data
\begin{eqnarray}\label{13}
\nonumber
W[h_{ij}',h_{ij}'',T]&=& \delta\Big( \partial_{i}\partial_{j}h^{\eps^i}_{ij}{}'' - \nabla^2 h{}^{\eps^i}{}'' \Big)
\; \delta \Big( \partial_{i}\partial_{j}h'_{ij} - \nabla^2 h{}' \Big)\ 
\int {\cal D}\epsilon^i
\no &&
 \int_{h_{ij}''{}^{\eps^i} \atop h_{ij}'} \mathcal{D}h_{ij} 
\;\Delta_{\rm FP}(h_{ij})\;
\exp \Bigg\{\frac{i}{4}\int d^3x \int_{0}^{T} dt \Big( \partial_\alpha h \partial^\alpha h
- \partial^{\alpha}h^{ij} \partial_\alpha h_{ij} 
\no \label{133}&&
+ \partial^k h^{ij} \partial_j h_{ik}  +\partial^k h^{ij} \partial_i h_{jk} - \partial_i h^{ij} \partial_j h - \partial^j h_{ij} \partial^i h \Big) 
 \Bigg\} .
\end{eqnarray}
The argument of the first $\delta$--function does not depend on $\eps^i$, since,
from the tranformation properties of $h_{ij}$ we have
$\partial_{i}\partial_{j}h^{\eps^i}_{ij}{}'' - \nabla^2 h{}^{\eps^i}{}''=
\partial_{i}\partial_{j}h''_{ij} - \nabla^2 h{}''$. Recalling also that the constant Fourier component of the boundary data vanishes because of the conditions at infinity, and absorbing a constant in the normalization factor, we can write
the $\delta$--functions simply as $\delta(h_0')\delta(h_0'')$. 
These two $\delta$--functions impose the linearized scalar constraint $h_0=0$ on the boundary data, and project on the solution of the quantum scalar constraint, namely the Wheeler--DeWitt equation.  As emphasized by Kucha\v{r} in \cite{Kuchar}, indeed, in the linear approximation the scalar constraint is not anymore a relation between momenta and configuration variables, but rather a condition on the configuration space: the solution of the linearized Wheeler--DeWitt equation are the wave functions with support on  $h_0=0$. 
Using this argument, in \cite{har} Hartle imposes these two $\delta$--functions on the ground state functional for Euclidean quantum gravity by hand. Here, instead, we obtain them as a result of the FP integration calculations.

Notice the two different ways in which the linearized temporal and spacial diffeomorphisms eliminate, respectively, the spin--zero $h_0$ and spin--one and spin--zero longitudinal components $h_i$ of the lineraized field $h_{ij}$. The physical states are independent from $h_i$, while they are concentrated on $h_0=0$.
Finally, we are left only with the physical spin--two field $h_{ij}^{\rm\scr TT}$.

The FP factor $\Delta_{\rm FP}$ is cubic in the fields \cite{teit2}. Since in the following we are interested in computing the gaussian approximation, we neglect it.  The gaussian integral over $h_{ij}$ is straightforward; we give the details of the integration in the appendix.  Writing
\eqa
H^{\rm \scriptscriptstyle TT}(\vec{p})&=& \overline{{h}^{\rm \scriptscriptstyle TT}{}''_{ij}(\vec{p})}\ {h}^{\rm \scriptscriptstyle TT}{}''_{ij}(\vec{p}) 
+ \overline{h^{\rm \scriptscriptstyle TT}{}'_{ij}(\vec{p})}\ h^{\rm \scriptscriptstyle TT}{}'_{ij}(\vec{p}), \no\label{H}
\tilde H^{\rm \scriptscriptstyle TT}(\vec{p})&=& \overline{ {h}^{\rm \scriptscriptstyle TT}{}''_{ij}(\vec{p})}\ h^{\rm \scriptscriptstyle TT}{}'_{ij}(\vec{p}) + \overline{h^{\rm \scriptscriptstyle TT}{}'_{ij}(\vec{p})}\ {h}^{\rm \scriptscriptstyle TT}{}''_{ij}(\vec{p}),
\neqa
the result of the integration is 
\begin{equation}
\label{19}
W[h_{ij}',h_{ij}'',T] = \mathcal{N} (T) \;
\delta(h_0')\; \delta (h_0'')\ 
e^{\frac{i}{4} {\textstyle
\int \frac{d^3p}{(2\pi)^3}  p 
\frac{ H^{\rm \scriptscriptstyle TT}(\vec p)\cos{p T} - \tilde H^{\rm \scriptscriptstyle TT}(\vec p)}{\sin{p T}}}},
\end{equation}
where the function $\mathcal{N} (T)$ is a normalization factor. 
This is the field propagation kernel of linearized GR. 

\subsection{Ground--state and graviton states}

We are now ready to read the vacuum state and the $n$--graviton states from the propagation kernel \Ref{19}. To do so, we expand \Ref{19} in power series of $e^{-ipT}$.
 To first order, we obtain  
\begin{equation}
\label{expansion}
W[h_{ij}',h_{ij}'',T] = \mathcal{N} (T) \;\delta(h_0')\; \delta (h_0'')\ e^{{-\frac{1}{4} 
\int \frac{d^3p}{(2\pi)^3}  p \;H^{\rm \scriptscriptstyle TT}(\vec p)}} 
\Bigg[1 + \frac{1}{2} \int \frac{d^3p}{(2\pi)^3} p \; 
\ H^{\rm \scriptscriptstyle TT} (\vec p) \;e^{-ipT} + \dots \Bigg]. 
\end{equation}
Using \Ref{vuoto}, the (non--normalized) vacuum state can be read from the zero'th order of \Ref{expansion}: we have
\begin{eqnarray}
\nonumber
\overline{\Psi_0[h''_{ij}]}\Psi_0[h'_{ij}] =\delta(h_0')\; \delta (h_0'')\  \exp \Big\{ -\frac{1}{4} 
\int \frac{d^3p}{(2\pi)^3}
p\;H^{\rm \scriptscriptstyle TT}(\vec p)  
\Big\}, \label{21}
\end{eqnarray}
and therefore
\begin{eqnarray}
\Psi_0 [h_{ij}] &=&  \delta(h_0)\  \exp \Big\{ -\frac{1}{4} 
\int \frac{d^3p}{(2\pi)^3} p \, \, h^{\rm \scriptscriptstyle TT}{}_{ij}(\vec{p})h^{\rm \scriptscriptstyle TT}{}_{ij}(-\vec{p})  \Big\}. \label{22}
\end{eqnarray}
This is in agreement with the literature \cite{Kuchar,har}.

The graviton states can be obtained from the analog of \Ref{ZYM2}, namely
$$
W[{h}_{ij}, h'_{ij}, T]= \frac{1}{n!}  \sum_{n=0}^\infty \;
\sum_{\epsilon_1...\epsilon_n}
\int \frac{d^3p_1}{(2\pi)^3}\ldots \frac{d^3p_n}{(2\pi)^3} 
\;e^{-i \sum_{m=1}^n E_m T} 
\;\overline{\psi_{p_1\epsilon_1,...,p_n\epsilon_n}[{h}_{ij}]} \;\psi_{p_1\epsilon_1,...,p_n\epsilon_n}[h'_{ij}].
$$
This expression can be matched with \Ref{expansion} to extract the $n$--graviton states.
The (non--normalized) wave functional of the one--graviton state with momentum $p$ and polarization $\epsilon$, for example, reads
\begin{equation}
\label{1grav}
\Psi_{p,\epsilon} [h_{ij}] =  \delta(h_0)\   \sqrt{p}\, \epsilon^{ij}  
\ h^{\rm \scriptscriptstyle TT}{}_{ij}(\vec{p})\ \Psi_0 [h_{ij}] 
\end{equation}
and so on.

\section{Newton potential from the propagation kernel}\label{newton}

In the presence of external static sources, the energy of the lowest energy state must be the Newton self--energy of the external source.  Therefore the Newton self--energy can be extracted from the field propagation kernel as its lowest Fourier component in $T$. 
This procedure has proven effective in YM theory \cite{rt}. In particular, in the abelian case, the external source can be taken to be static, and the lowest energy state is characterized by the Coulomb self--energy. In the non--abelian case, on the other hand, the external sources cannot be static, reflecting the exchange of colour charges between the external sources and the system.
Inserting an external source in GR is a more delicate procedure, because the Newton potential emerges not only in the non--relativistic limit, as the Coulomb potential in YM theory, but also in the low gravity limit. This means that we can follow the same procedure one uses in YM theory,
but we expect for the Newton potential to emerge only in the linear approximation. 

To introduce an external source $\mathcal{J}_{\mu\nu}(\vec x, t)$, we consider the lagrangian
\equ\label{Lmatter}
\mathcal{L}_m[g_{\mu\nu}; \mathcal{J}_{\mu\nu}]=
{\mathcal L}[g_{\mu\nu}]-g^{\mu\nu}\mathcal{J}_{\mu\nu},
\nequ
The source is the densitized matter energy--momentum tensor 
$\mathcal{J}_{\mu\nu}= {\sqrt{-g}}\,{T}_{\mu\nu}$.
In the linear approximation around Minkowski space, the lagrangian \Ref{Lmatter} becomes
\equ\label{LmatterLin}
\mathcal{L}_m[h_{\mu\nu}; \mathcal{J}_{\mu\nu}]=\mathcal{L} [h_{\mu\nu}]
-\eta^{\mu\nu}\mathcal{J}_{\mu\nu} +  h^{\mu\nu}\mathcal{J}_{\mu\nu}.\label{74}
\end{equation}
We characterize a static external source with the condition that only the component 
$\mathcal{J}_{00}=\rho$ be different from zero. $\rho(\vec x, t)$ is thus the energy density
of the source.  Covariance is broken since the source itself defines a 
preferred frame. The source term in \Ref{LmatterLin} reads then 
$\rho + h^{00}\rho.$
Note that the conservation low $\n_\mu  T^\mu{}_\nu=0$ at first order reads $\p_0 \rho=0$, consistently with the
fact that the source is static.  In the temporal gauge $h_{0\mu}=0$ the Lagrangian becomes
\begin{equation}
\mathcal{L}_m[h_{ij}, h_{0\mu}=0; \mathcal{J}_{\mu\nu}]=\mathcal{L}[h_{ij}]+\rho.\label{75}
\end{equation}
Notice that we seem to lose the coupling between the gravitational field and the external source. This is analogous to what happens in YM, where the coupling term $A_0 \rho$ to an external static source is killed by the temporal gauge $A_0=0$. This seems to prevent us from coupling external sources to the field in the temporal gauge. But the coupling can nonetheless be obtained, because part of the gauge degrees of freedom turn out to describe the source \cite{rt}.  
Below we show how this happens.
Indeed, staring from  lagrangian \Ref{75} and following the same steps as in the previous sections we arrive at the following expression for the propagation kernel,
\begin{eqnarray}
W[h_{ij}',h_{ij}'',T]&=& \mathcal{N}(T) \; \delta
\left(\partial_i\partial_jh''_{ij}-\n^2h{}'' + \rho\right) 
\delta \left(\partial_i\partial_jh'_{ij}-\n^2h{}' + \rho\right) \exp\left\{ i\,T\int \rho \, d^3x  \right\} \cdot \no
&&\cdot\exp\Bigg\{\frac{i}{4} \int \frac{d^3p}{(2\pi)^3} \Bigg[  p^2 T \ \frac{ 2 H_{ijkl}(\vec p) + \tilde H_{ijkl}(\vec{p})}{12} \ D_{ij}D_{kl} +\no
&&+  \frac{ H_{ijkl}(\vec p) - \tilde H_{ijkl}(\vec{p})}{2p^2T} \Big( -p^2\;D_{ij}D_{kl} - 2 p_i p_j D_{kl}- 2 p_k p_l D_{ij}  \Big) + \no \label{77}
&& +p\;\frac{ H_{ijkl} (\vec{p})\cos{p T} - \tilde H_{ijkl}(\vec{p})}{2\sin{p T}} \Big( D_{ik}D_{jl} + D_{il}D_{jk} -  D_{ij} D_{kl}  \Big) \Bigg] \Bigg\},
\end{eqnarray}
where
\begin{eqnarray}H_{ijkl}(\vec{x},\vec{y})&=& h''_{ij}(\vec{y})h''_{kl}(\vec{x}) + h'_{ij}(\vec{y})h'_{kl}(\vec{x}), \no
\tilde H_{ijkl}(\vec{x},\vec{y})&=& h''_{ij}(\vec{y})h'_{kl}(\vec{x}) + h'_{ij}(\vec{y})h''_{kl}(\vec{x}).
\label{HH}
\end{eqnarray}
As before, the integrals over the $\eps^0$'s implement the Wheeler--DeWitt constraint, but 
in presence of matter this is now $\partial_i\partial_j h''_{ij}-\n^2h'' + \rho$, or, in momentum
space, $2 p^2 h_0 - \rho(\vec p)$. 
 Using these constraints we get
\begin{equation}
\label{79}
W[h_{ij}',h_{ij}'',T]= \mathcal{N}(T) \; e^{-i \left(E_0 - m \right) T}\;
\exp\Bigg\{\frac{i}{4} \int \frac{d^3p}{(2\pi)^3}   
\Bigg[  p\; \frac{ H^{TT}_{ijkl} (\vec{p})\cos{p T} - \tilde H^{TT}_{ijkl}(\vec{p})}{\sin{p T}} \Bigg]\Bigg\},
\end{equation}
where
\equ\label{EN}
E_0=-\frac{1}{32\pi}\int d^3x \int d^3y 
\frac{ \rho(\vec x)\rho(\vec y)}{|\vec x -\vec y|}, \qquad \qquad m = \int \rho (\vec x) \, d^3 x.
\nequ
Therefore, the ground state is now weighted by the classical expression for the Newtonian
self--energy of the external source and by the its rest mass.  In the presence of sources, the field propagation nicely codes the Newtonian interaction.

\section{Summary}

We have studied the field propagation kernel of general relativity in the temporal gauge, 
$W[g'_{ij}, g''_{ij}, T]$. This is given in equation \Ref{6}.  It can be interpreted as a 3--geometry to 3--geometry transition amplitude, or as a projector on the solutions of the quantum constraints.   It is independent from $T$ when $g'_{ij}$ and $g''_{ij}$ are defined on compact spaces; it depends on the asymptotic time if they are asymptotically flat.  When $g'_{ij}$ and $g''_{ij}$ are close to $\delta_{ij}$, it can be computed explicitly to first order: 
the resulting expression is given in \Ref{19}. It is independent from the spin--one and spin--zero longitudinal components $h_i$ of the linearized field, and concentrated on the spin--zero component $h_0=0$ values. From its form, the linearized vacuum, given in \Ref{22}, and $n$--graviton states, as in \Ref{1grav}, follow immediately.  By coupling a static source to the field, we can extract  the Newtonian self--energy \Ref{EN} from the linearized expression of the kernel.

These results are based on an accurate implementation of the FP procedure. The projection on the solutions of the constraints is implemented by the integration over the FP gauge parameters. This happens in a more subtle way than in YM theory, since temporal diffeomorphisms affect the boundary of the action. 

We have disregarded ultraviolet divergences.  There are tentative background--independent definitions of quantum gravity that define 3--geometry to 3--geometry transition amplitudes free from ultraviolet divergences.  In order for these to have an acceptable low energy limit, they should reduce to  the propagator kernel computed here, for values of their arguments close to flat space. 

Finally, showing that the background--independent definitions of quantum gravity lead to the correct Newtonian interaction at low energy has proven so far surprisingly elusive.  Here we have 
shown how to extract this interaction from the field propagation kernel, by adding an external source.   

\vskip1cm

MT thanks the MIUR (Italian Minister of Instruction University and Research) and INFN (Italian National Institute for Nuclear Physics) for partial support.   CR thanks the physics department of the University of Rome ``La Sapienza" and FM, MT and SS thank the CPT  for hospitality during the preparation of this work. 

\vskip1cm

\appendix
\section*{Appendix A: Hamilton function of linearized GR}

In this appendix we evaluate the Hamilton function of linearized GR.  This is the value
of the gravitational action on a classical solution with given boundary values, expressed as a function of these boundary values \cite{r}. This will be used in the next appendix, to obtain the expression (\ref{19}) of the propagation kernel. To simplify the calculations, we work with Euclidean signature, using analytic continuation to switch back to Lorentzian signature at the end of the calculations. In this appendix, $x^0$ is the Euclidean time, defined by
$x^0=it$.

The quadratic part of the action of the linearized field, in the temporal gauge is
\begin{equation}
\label{a1}
S=\frac{1}{4}\int d^4 x \Big[ \partial^k h^{ij} \partial_j h_{ik} + \partial^k h^{ij} \partial_i h_{jk} -
 \partial^\alpha h^{ij} \partial_\alpha h_{ij} - \partial_i h^{ij} \partial_j h - 
 \partial^j h_{kj} \partial^k h + \partial^\alpha h \partial_\alpha h \Big].
\end{equation}
In the following calculations the linear term in \Ref{actionLin} plays no role. The equations of motion obtained by varying this action are
\begin{equation}
\label{a2}
\square \,h_{ij} - \partial_i \partial^m h_{mj} - \partial_j \partial^m h_{im} + \partial_i \partial_j h 
+ \delta_{ij} \partial^m \partial^n h_{mn} - \delta_{ij} \square\, h  = 0.
\end{equation}
We are interested in evaluating the action on the classical solution with boundary data $h'_{ij}$ and $h''_{ij}$.
Using the field equations and the boundary data, the action can be written as
\begin{eqnarray}
\label{a13}
S_{cl} = - \frac{1}{4} (\delta_{a}^k \delta_{b}^l - \delta_{ab} \delta^{kl}) 
\int d^3x \Bigg[ {h''}^{ab} (\vec{x}) \partial_0^x h_{kl}(x)\Big|_{x^0=T} - 
{h'}^{ab} (\vec{x}) \partial_0^x h_{kl}(x)\Big|_{x^0=0}\Bigg],
\end{eqnarray}
where the $h_{ij}(\vec x, t)$ is the classical solution interpolating between $h'_{ij}$ and $h''_{ij}$. This classical solutions can be constructed by means of a Green function $G_{ijkl}(x,y)$, defined by
\begin{eqnarray}
\label{a3}
&&\square \, G_{ijkl}(x,y) - \partial_i \partial^m G_{mjkl}(x,y) - \partial_j \partial^m G_{imkl}(x,y) + \partial_i \partial_j {G^m}_{mkl}(x,y)+\\ \nonumber
&& + \delta_{ij} \partial^m \partial^n G_{mnkl}(x,y) - \delta_{ij} \square\, {G^m}_{mkl} (x,y) = \frac{1}{2}\left( \delta_{ik}\delta_{jl}+\delta_{il}\delta_{jk}\right)\delta^{(4)}(x-y),
\end{eqnarray}
with boundary conditions $G_{ijkl} (\vec{x},0;y)=G_{ijkl} (\vec{x},T;y) = 0$.
To find it, we Fourier transform in the spacial components,
$$ 
G_{ijkl}(x,y)=\int\frac{d^3p}{(2\pi)^3} \tilde G_{ijkl}(\vec{p};x^0,y^0) e^{-i\vec{p}(\vec{x}-\vec{y})};
$$
then decompose $\tilde G_{ijkl}$ into covariant tensors and match the corresponding
terms in \Ref{a3}. We obtain
\eqa\nonumber
&& \tilde G_{ijkl} (\vec{p}; x^0, y^0) =  A({p^2}, x^0, y^0)\; p_i p_j p_k p_l +
 \\ \nonumber
&&+  B({p^2}, x^0, y^0) 
 \Big[ -p_i p_j p_k p_l + p^2 \big( p_i p_k \delta_{jl}  + 
 p_j p_l \delta_{ik}+ p_i p_l \delta_{jk} +p_j p_k \delta_{il} 
 -p_i p_j \delta_{kl} -p_k p_l \delta_{ij}\big) \Big]+ \\ \nonumber
&& +C({p^2}, x^0, y^0)  
\Big[ p^4 \left( \delta_{ik}\delta_{jl} +  \delta_{il}\delta_{jk} - \delta_{ij}\delta_{kl} \right) 
+ \\ 
&& - p^2 \big( p_i p_k \delta_{jl} + p_j p_l \delta_{ik} + 
 p_i p_l \delta_{jk} +p_j p_k \delta_{il} -p_i p_j \delta_{kl} 
 -p_k p_l \delta_{ij}\big) + p_i p_j p_k p_l \Big],\label{a11}
\end{eqnarray}
where the functions $A, B, C$ are defined as follows:
\eqa 
A({p^2}, x^0, y^0)&=&- \frac{{y^0}^3+2T^2 y^0-3{y^0}^2T + \left(y^0 -T \right){x^0}^2}{12p^2T}x^0 -
 \frac{\left(x^0-y^0\right)^3}{12p^2} \theta(x^0-y^0);
\no
B({p^2}, x^0, y^0)&=& \left(y^0 -T \right) \frac{x^0}{2p^4T}    +  \frac{\left(x^0-y^0\right)}{2p^4} \theta(x^0-y^0);
\no\nonumber
C({p^2}, x^0, y^0)&=&\frac{\sinh p x^0 \sinh{(p y^0-pT)}}{2p^5\sinh{p T}}\theta(y^0-x^0) 
+\frac{\sinh(p x^0-pT) \sinh{p y^0}}{2p^5\sinh{p T}}\theta(x^0-y^0).
\neqa
The Green function allows us to write the solution of the field equations with given boundary data
\begin{equation}
\label{a5}
h_{kl} (y) = \left( \delta_{im}\delta_{jn} - \delta_{ij}\delta_{mn} \right)
\!\int \!\!d^3x \Bigg\{ h''_{mn} (\vec{x}) \partial^x_0 G_{ijkl}(x,y)\Bigg|_{x^0=T} -
 h'_{mn} (\vec{x}) \partial^x_0 G_{ijkl}(x,y)\Bigg|_{x^0=0} \Bigg\}.
\end{equation} 
By inserting \Ref{a11} into \Ref{a5} and this into (\ref{a13}), we obtain the Hamilton function of unconstrained linearized GR
\begin{eqnarray}
\nonumber
S_{cl}[h_{ij}',h_{ij}'',T]\!\!\!&=&\!\!\! -\frac{1}{4} \int \frac{d^3p}{(2\pi)^3}  
\Bigg\{  -p^2 T\; \frac{ 2 H_{ijkl}(\vec{p}) + \tilde H_{ijkl}(\vec{p})}{12}  \ D_{ij}D_{kl} + 
\frac{ H_{ijkl}(\vec{p}) - \tilde H_{ijkl}(\vec{p})}{2p^2T} 
\cdot \no &&\cdot
\Bigg[ -p^2 \ D_{ij}D_{kl} + p_i p_k D_{jl} + p_j p_l D_{ik} + p_i p_l D_{jk} + p_j p_k D_{il} - 2 p_i p_j D_{kl} 
- 2 p_k p_l D_{ij}  \Bigg] 
 \no 
&&+p\; \frac{ H_{ijkl} (\vec{p})\cosh{p T} - \tilde H_{ijkl}(\vec{p})}{2\sinh{p T}} \Bigg[ D_{ik}D_{jl} + D_{il}D_{jk} -  D_{ij} D_{kl}  \Bigg] \Bigg\}.\label{a14}
\end{eqnarray}
where $D_{ij}$ and $H_{ijkl}$ are given in \Ref{D} and \Ref{HH}.

\section*{Appendix B: Evaluation of the propagation kernel}

Here we illustrate the derivation of  \Ref{19} from  \Ref{13}. To perform the gaussian integral over $h_{ij}$, we write the field as the solution of the classical equations of motion  $h_{ij}^0$ with the given boundary data, plus a fluctuation $\zeta_{ij}$. The boundary data are then 
\begin{equation}\zeta_{ij}(0)=\zeta_{ij}(T)=0. 
\end{equation}Linear terms in $\zeta_{ij}$ vanish because of the equations of motion, the integration over $\zeta_{ij}$ yields a normalization factor $\mathcal{N}(T)$ depending of $T$, but independent of the boundary data.   The remaining exponential, independent of the perturbation field, is the Hamilton function, computed above. Using this, we obtain
for the Euclidean propagation kernel
\eqa
W_E[h_{ij}',h_{ij}'',T]\!\!\! &=&\!\!\! \mathcal{N} (T) \;
\delta\left( \partial_{i}\partial_{j}h{}''_{ij} - \nabla^2 h{}'' \right)
\; \delta \left( \partial_{i}\partial_{j}h'_{ij} - \nabla^2 h' \right)\cdot \no
&& \cdot \int {\cal D}\eps^i(T,\vec{x})\;
\exp\Bigg\{-\frac{1}{4}\int \frac{d^3p}{(2\pi)^3}\int d^3x \int d^3y \;e^{-i\vec p (\vec y-\vec x)} \cdot \no
&& \cdot \Bigg[  -p^2\;\frac{ 2 H^\epsilon_{ijkl}(\vec{x},\vec{y}) + \tilde H^\epsilon_{ijkl}(\vec{x},\vec{y})}{12} T \; D_{ij}D_{kl} +  
\no && + \frac{ H^\epsilon_{ijkl}(\vec{x},\vec{y}) - \tilde H^\epsilon_{ijkl}(\vec{x},\vec{y})}{2p^2T} 
\Big( -p^2D_{ij}D_{kl} + \no && + p_i p_k D_{jl} + p_j p_l D_{ik} + p_i p_l D_{jk} + p_j p_k D_{il} - 2 p_i p_j D_{kl} - 2 p_k p_l D_{ij}  \Big) + \no \label{14}
&& +p\;\frac{ H^\epsilon_{ijkl} (\vec{x},\vec{y})\cosh{p T} - \tilde H^\epsilon_{ijkl}(\vec{x},\vec{y})}{2\sinh{p T}} \Big( D_{ik}D_{jl} + D_{il}D_{jk} -  D_{ij} D_{kl}  \Big) \Bigg] \Bigg\}.
\neqa
Here the apex $\eps$ indicates a spatial diffeomorphism $\eps^i$ only on the field $h''_{ij}$, 
\eqa\label{15}
H^\epsilon_{ijkl}(\vec{x},\vec{y})&=& {h}^{\eps^i}{}''_{ij}(\vec{y}){h}^{\eps^i}{}''_{kl}(\vec{x}) + h'_{ij}(\vec{y})h'_{kl}(\vec{x}), \\ \label{17}
\tilde H^\epsilon_{ijkl}(\vec{x},\vec{y})&=& {h}^{\eps^i}{}''_{ij}(\vec{y})h'_{kl}(\vec{x}) + h'_{ij}(\vec{y}){h}^{\eps^i}{}''_{kl}(\vec{x}).
\end{eqnarray}
The term in the last line of the expression \Ref{14} has the structure of the
propagation kernel for a harmonic oscillator (see for instance \cite{r}).

The next step is to consider the integration over the spatial diffeomorphisms; 
the integrals can be performed separately for the longitudinal and the transversal parts\footnote{We have
the usual decomposition of a 3--vector, $\eps^i = \eps_{\rm \scr L}^i + \eps_{\rm \scr T}^i =
\frac{1}{\n^2}\p^i\p_j\eps^j + \eps^i-\frac{1}{\n^2}\p^i\p_j\eps^j$.} of $\eps^i$.
When we perform the integration over the transversal spatial diffeomorphisms,
the result we obtain is that only the term with the structure of a harmonic oscillator survives.
On the other hand, the integration over the longitudinal part $\p_i\eps^i$ gives again
an implementation of the scalar constraint, in the form
\equ\label{redund}
\delta\left( \partial_i\partial_j h''_{ij}-\n^2h''-\partial_i\partial_j h'_{ij}+\n^2h'\right).
\nequ
The redundance of the integration over the $\eps^0$'s and the $\p_i\eps^i$ can be understand as follows:
in the hamiltonian formalism, the scalar constraint is proportional to the lapse function $N$. In the
``Lapse=1, Shift=0'' gauge we are working it, the transformation of the lapse under a diffeomorphism
is given by (see for instance \cite{teit2}) 
$N^\eps = 1 + \p_0 \eps^0 + \p_i\eps^i$. Therefore, both the integration over $\eps^0$
and over $\p_i \eps^i$ provide variation of the lapse and consequentely an implementation of the
scalar contraint.

In the case of matter, the redundant implementation of the scalar constraint \Ref{redund},
coming from the integration over the longitudinal spatial diffeomorphisms, is consistent
with the other $\delta$--functions, because the source is static and thus $\rho(T)=\rho(0)$.

The final expression can be extended to Lorentzian signature by means of the analytic continuation $T\to iT$, and we obtain
\eqa
W[h_{ij}',h_{ij}'',T] &=& \mathcal{N} (T) \;
\delta\left( \partial_{i}\partial_{j}h{}''_{ij} - \nabla^2 h{}'' \right)
\; \delta \left( \partial_{i}\partial_{j}h'_{ij} - \nabla^2 h' \right)\cdot \no
&& \cdot\exp\Bigg\{\frac{i}{4} \int d^3x \int d^3y
\int \frac{d^3p}{(2\pi)^3} e^{-i\vec{p}(\vec{y}-\vec{x})} p 
\frac{ H_{ijkl}(\vec{x},\vec{y})\cos{p T} - \tilde H_{ijkl}(\vec{x},\vec{y})}{2\sin{p T}} \cdot
\no\label{1000}
&& \cdot \left( D_{ik}D_{jl} + D_{il}D_{jk} -  D_{ij}D_{kl}  \right) \Bigg\},
\neqa
which coincides with \Ref{19}, once the spacial indices are contracted.

\section*{Appendix C: Linearized Einstein equations in the temporal gauge}
For completeness, we write here the linearized Einstein equations in the temporal gauge $h_{0\mu}=0$, extensively utilized in this paper. To do so it is convenient to work in the Fourier space with the Poincar\'e--irreducible components introduced in (\ref{poincare}). These can be obtained from the field as follows
\begin{eqnarray}
\nonumber
h_0 = \frac{1}{2} h_{kl} D_{kl},\ &\qquad& h_{\rm \scr L} = h_{kl} \frac{p_k p_l}{p^2},  \\ \label{e5}
h^{\rm \scr T}_i = h_{kl} \frac{p_k}{p} D_{il}, &\qquad& h^{\rm \scriptscriptstyle TT}_{ij} = h_{kl} \left( D_{ik} D_{jl} - \frac{1}{2} D_{ij} D_{kl} \right).
\end{eqnarray}
We consider the case of a dust distribution with energy density $\rho(\vec x)$. So the only non--vanishing component of the energy--momentum tensor is $T_{00}=\rho$. As pointed out in section (\ref{newton}), the source has to be static in the linearized case because of the continuity equation. We will found this result also in the study of the Einstein equations; this is not surprising because the continuity equation comes from the Bianchi identities, which are properties of the Einstein equations themselves. The vacuum case is easily reconstructed setting $\rho=0$.

The linearized equations for a dust distribution are
\begin{equation}
\label{e8}
\partial_\alpha \partial^\alpha h_{\mu\nu} - \partial_\mu \partial^\alpha h_{\alpha\nu} - \partial_\nu \partial^\alpha h_{\alpha\mu} + \partial_\mu \partial_\nu h^{\alpha}{}_{\alpha}=\frac{\rho}{2} \delta_{\mu\nu}.
\end{equation}
They can be written in the temporal gauge as follows 
\begin{eqnarray}
\label{e18}
&&\ddot h^i{}_i=\frac{\rho}{2}, \\ \label{e19}
&&-\partial^j \dot h_{ij} + \partial_i \dot h^k{}_k =0, \\ \label{e20}
&&\square h_{ij} - \partial_i \partial^k h_{kj} - \partial_j \partial^k h_{ji} + \partial_i \partial_j h^{k}{}_{k}=\frac{\rho}{2}\delta_{ij},
\end{eqnarray}
where the dot indicates the derivative respect to the zero component and  $\square= \partial_\alpha \partial^\alpha$.
From the time--equations (\ref{e18},\ref{e19}) we obtain the constraints in terms of the quantities defined in (\ref{e5})
\begin{equation}
\label{e12}
\dot h_0=0, \qquad \dot h^{\rm \scr T}_i=0, \qquad \ddot h_{\rm \scr L}=\frac{\rho}{2}.
\end{equation}
On the other hand, imposing the constraints on the motion equations (\ref{e20}) we have
\begin{equation}
\label{e23}
\square h^{\rm \scriptscriptstyle TT}_{ij}=0, \qquad 2 p^2 h_0=\rho.
\end{equation}
The first is a wave equation for the traceless transverse components, which are the only physical degrees of freedom. The second is the Wheeler--deWitt equation in the Fourier space, as can be easily seen from the definition of $h_0$ in (\ref{e5}). As it was previously said, the first of the constrains (\ref{e12}) together with the Wheeler--deWitt equation imposes the source to be static.

In terms of the Poincar\'e-irreducible field components, the quadratic term of the Lagrangian can be written as 
\begin{equation}
\label{e16}
4 \mathcal{L}^{(2)} = h^{\rm \scriptscriptstyle TT}_{ij} \square h^{\rm \scriptscriptstyle TT}_{ij} + 2 h_0 \square h_0 + 4 \dot h_0 \dot h_{\rm \scr L} -2 \dot h^{\rm \scr T}_i \dot h^{\rm \scr T}_i. 
\end{equation}

\end{document}